\documentclass[iop,apjl]{emulateapj}
\usepackage{apjfonts,graphicx}
\shorttitle{The velocity dispersion of Pal 14}
\shortauthors{Sollima et al.}
\begin{document}
\title{A Monte Carlo analysis of the velocity
dispersion of the globular cluster Palomar 14}
\author{A. Sollima\altaffilmark{1}, C. Nipoti\altaffilmark{2}, 
A. Mastrobuono Battisti\altaffilmark{3}, M. Montuori\altaffilmark{3}, R.
Capuzzo-Dolcetta\altaffilmark{3}}
%
%\email{asollima@iac.es}
%
%
\altaffiltext{1}{INAF Osservatorio Astronomico di Padova, 
               vicolo dell'Osservatorio 5, I-35122 Padova, Italy}
\altaffiltext{2}{Dipartimento di Astronomia, Universit\`a di Bologna, I-40127
Bologna, Italy}
\altaffiltext{3}{Dipartimento di Fisica, Universit\`a di Roma La Sapienza,
Piazzale Aldo Moro 2, I-00185 Rome, Italy}
\begin{abstract}
We present the results of a detailed analysis of the projected
velocity dispersion of the globular cluster Palomar 14 performed using
recent high-resolution spectroscopic data and extensive Monte Carlo
simulations. The comparison between the data and a set of dynamical
models (differing in fraction of binaries, degree of anisotropy,
mass-to-light ratio $M/L$, cluster orbit and theory of gravity) shows
that the observed velocity dispersion of this stellar system is well
reproduced by Newtonian models with a fraction of binaries
$f_{b}<30\%$ and a $M/L$ compatible with the predictions of stellar
evolution models. Instead, models computed with a large fraction of
binaries systematically overestimate the cluster velocity
dispersion. We also show that, across the parameter space sampled
  by our simulations, models based on the Modified Newtonian Dynamics
  theory can be reconciled with observations only assuming values of
  $M/L$ lower than those predicted by stellar evolution models under
  standard assumptions.
\end{abstract}
\section{Introduction}
\label{intro_sec}

Palomar 14 is one of the least luminous globular clusters (GCs) of the
Milky Way ($M_{V}=-4.95\pm0.12$) and it is located in the outer halo
of the Milky Way at a distance from the Sun $d\sim71~kpc$ (Sollima et
al. 2011, hereafter S11). These characteristics make this object
particularly interesting from a kinematical point of view: both its
internal and external accelerations are weaker than the characteristic
acceleration of the Modified Newtonian Dynamics (MOND; Milgrom 1983)
$a_{0}\simeq1.2\times 10^{-10} m/s^{2}$ and its low binding energy
makes this stellar system prone to a significant tidal stress.

In recent years, this cluster has been the object of many
investigations focused on its peculiar structure and kinematics.  Pal
14 has been indeed indicated as one of the best candidates to test
MOND (Baumgardt et al. 2005; Sollima \& Nipoti 2010; Haghi et
al. 2009, 2011) because the global projected velocity dispersions
predicted by the classical Newtonian theory and MOND differ
significantly for this stellar system.  Unfortunately, the distance
and mass of Pal 14 imply that only a bunch of Red Giant Branch (RGB) stars within the
half-mass radius appear brighter than the limiting magnitude
($V\sim20$) of high-resolution spectrographs mounted on 8m-class
telescopes. For this reason, the only high-resolution spectroscopic
analysis available for Pal 14 (Jordi et al. 2009; hereafter J09)
derived accurate radial velocities for only 17 member stars.  J09
compared the projected velocity dispersion of the system with the
outcome of a set of N-body simulations, reporting that the expected
velocity dispersion in MOND is more than three times higher than the
observed value. They concluded that the measured velocity dispersion
of Pal 14 represents a problem for MOND. Gentile et al. (2010), using
a Kolmogorov-Smirnov test on the observed and predicted MOND
distributions of velocities, claimed that the confidence level
achievable using the small sample of stars used by J09 does not allow
to rule out MOND.

On the other hand, Kupper \& Kroupa (2010) used a large set of N-body
realizations of Pal 14 in the framework of the classical Newtonian
dynamics including the effect of a variable fraction of binaries and
found that a fraction of binaries $f_{b}>10\%$ would be incompatible
with the observed velocity dispersion. They argued that Newtonian
gravity is challenged by the observed kinematics of Pal 14, unless the
cluster hosts an unusually small number of binaries. However, a larger
binary fraction is expected to be present in this loose cluster both
from theoretical arguments (Kroupa 1995) and from considerations on
the observed fraction of Blue Straggler Stars (Beccari et
al. 2011). In this case, according to Kupper \& Kroupa (2010), the
observational evidence would suggest a gravitational force {\it
  weaker} than the Newtonian one in the low-acceleration regime
(i.e. in the opposite direction of what MOND predicts). 

Recent attempts to model the kinematics of Pal 14 with N-body
simulations have also revealed difficulties in reproducing the
present-day mass-function and structure of this stellar system
starting from a standard Initial Mass Function (IMF; Zoonozi et
al. 2011). Therefore, peculiar starting conditions (like flattened IMF
and/or primordial mass-segregation) could characterize this cluster.
The situation recently became even more complicated: a pair of tidal
tails surrounding the cluster has been discovered (S11), suggesting
that the Galactic tidal field is important and questioning the
validity of models that treat the cluster as isolated.

In this paper we try to shed light on the controversial interpretation
of the kinematics of Pal 14 by presenting a statistical analysis of
its projected velocity dispersion based on a Monte Carlo approach. We
used the sample of radial velocities of J09 and a set of N-body
simulations in both Newtonian gravity and MOND, investigating the
effect of different assumptions on $M/L$, binary fraction, degree of
anisotropy and cluster orbit.

\section{Radial velocities}
\label{data_sec}

For the present analysis we used the sample of radial velocities
obtained by J09. It consists of 27 stars observed in the innermost 4
arcminutes of Pal 14 selected along the cluster Red Giant Branch.  It
has been constructed from high-resolution ($45,000<R<60,000$) spectra
collected with two different spectrographs: UVES@VLT and HIRES@KeckI.
A detailed description of the reduction procedure and of the radial
velocity measure can be found in J09. The high-resolution allows to
obtain very accurate radial velocities with errors of $\sim0.3~km/s$.
A number of outliers (field stars) can be easily identified at
velocities $\Delta v\equiv\vert v-\overline{v}\vert>5~km/s$. The
bona-fide cluster members turn out to constitute a sample of 16 stars
plus one star (\#15) lying at $\Delta v=2.35~km/s$ ($\sim 4\sigma$) so
that it is not clear if it is a true cluster member, a binary star or
a field outlier. In the following analysis we will always consider the two
samples defined including and excluding this star.  The error weighted averages of the bona-fide cluster members are
$\overline{v}=72.34\pm0.05~km/s$ (without star \#15) and
$\overline{v}=72.31\pm0.05~km/s$ (with star \#15).  To calculate the
velocity dispersion we searched for the value of $\sigma_{v}$ that
maximizes the logarithm of the probability density
\begin{eqnarray}
\label{sig_eq}
\ell&=&\sum_{i} \ln \int_{-\infty}^{+\infty}\frac{\exp\left[-\frac{(v'-\overline{v})^{2}}{2\sigma_{v}^{2}}-
\frac{(v_{i}-v')^2}{2\delta_{i}^{2}}\right]}
{2\pi \sigma_{v} \delta_{i}} dv' \nonumber\\
&=&-\frac{1}{2}\sum_{i}\left(
\frac{(v_{i}-\overline{v})^2}{\sigma_{v}^2+\delta_{i}^{2}}
+\ln [2\pi(\sigma_{v}^2+\delta_{i}^{2})]\right),
\end{eqnarray}
where $v_{i}$ and $\delta_{i}$ are the velocity of the $i$-th star and
its associated uncertainty (see Pryor \& Meylan 1993).  The
so-calculated velocity dispersions turn out to be
$\sigma_{v}=0.39_{-0.09}^{+0.14}~km/s$ (without star \#15) and
$\sigma_{v}=0.66_{-0.12}^{+0.19}~km/s$ (with star \#15).

\section{Models}
\label{mod_sec}

The models adopted in this paper are based on a set of N-body
simulations performed in the framework of both the Newtonian and MOND
theories of gravity considering the cluster immersed in the
gravitational field of the Milky Way. As we will show in
Sect. \ref{nbody_sec}, the effect of the external field is an
essential ingredient for a proper comparison of the velocity
dispersion of this low-mass cluster with observations, in particular
when the MOND theory is considered. It is however instructive to show
also the predictions of a set of isolated analytical models to provide
the necessary reference to quantify the effect of the external field
and to investigate the effect of different degrees of anisotropy.  In
the following sections we will describe the complete set of models and
simulations used in this paper.

\subsection{Analytical models}
\label{analyt_sec}

\begin{figure*}
\epsscale{1.}
\plotone{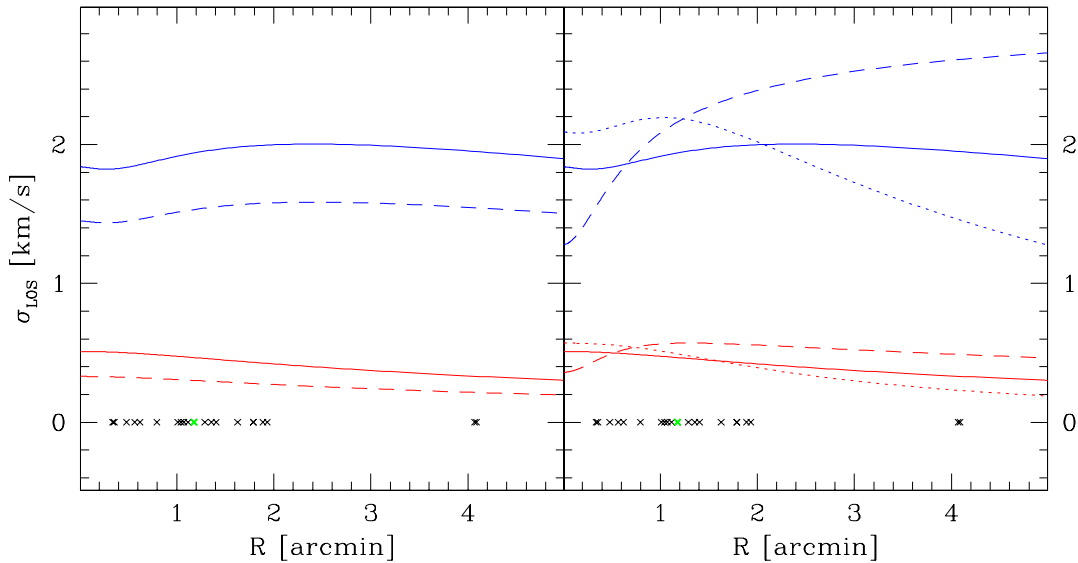}
\caption{Line-of-sight velocity dispersion profiles predicted by
  Newtonian (black lines; colored red in the online version) and MOND 
  (grey lines; blue in the online version) models of Pal 14,
  assuming the cluster isolated. In the left panel isotropic models
  with different $M/L$ ratios ($M/L$=1.885: solid lines; $M/L$=0.747:
  dashed lines) are shown. In the right panel models with $M/L$=1.885
  and different degrees of anisotropy (isotropic: solid lines; pure
  tangential: dashed lines; extremely radial: dotted lines) are
  shown. The radial locations of the target stars are marked at the
  bottom of both panels with black crosses. The thick cross (colored green in
  the online version) marks the location of star \# 15.}
\label{sig}
\end{figure*}

Under the hypothesis that the cluster is isolated and spherically
symmetric,  simple models can be  constructed by solving the Jeans
equation 
\begin{equation}
\label{jeans_eq}
\frac{1}{\rho}\frac{d \sigma_{r}^{2} \rho}{d r}+2 \beta
\frac{\sigma_{r}^{2}}{r}=-\frac{d\Phi}{d r},
\end{equation}
where $\rho$ is the density, $\Phi$ is the gravitational potential,
\begin{equation}
\label{beta_eq}
\beta(r)=1-\frac{\sigma_{t}^{2}(r)}{2\sigma_{r}^{2}(r)}
\end{equation} 
is the anisotropy parameter, and $\sigma_{r}$ and $\sigma_{t}$ are,
respectively, the radial and tangential components of the velocity
dispersion tensor at the radius $r$. If the system is
self-gravitating, $\rho$ and $\Phi$ are related in Newtonian gravity
by the Poisson equation
\begin{equation}
\label{pois_eq}
\nabla^{2}\Phi=4\pi G \rho,
\end{equation}
while in MOND by the corresponding modified field equation (Bekenstein
\& Milgrom 1984)
\begin{equation}
\label{mond_eq}
\nabla \cdot \left[ \mu\left( \frac{\|\nabla\Phi\|}{a_{0}}\right)\nabla\Phi\right]=4\pi G \rho,
\end{equation}
where $\mu(x)$ is the so-called "interpolating function" such that
$\mu(x)\sim x$ at $x\ll 1$ (the so-called "deep-MOND" regime) and
$\mu(x)\sim1$ at $x\gg 1$ (the Newtonian regime). In the following we
use the "simple" interpolating function $\mu(x)=x/(1+x)$ (Famaey \&
Binney 2005).  For finite-mass isolated systems the boundary
conditions of equations~(\ref{pois_eq}) and (\ref{mond_eq}) are $|
\nabla\Phi |\to 0$ for $|{\bf r}|\to \infty$, where ${\bf r}$ is the
position vector.

For given density distribution and $\beta(r)$,
equation~(\ref{jeans_eq}) can be solved to obtain $\sigma_{r}$.  The
tangential component of the velocity dispersion $\sigma_{t}$ is then
derived from eq.~(\ref{beta_eq}) and the line-of-sight (LOS) velocity
dispersion at any given projected distance $R$ from the center is
given by
\begin{equation}
\label{proj_eq}
\sigma_{LOS}^{2}(R)=\frac{2}{\Sigma(R)}\int_{R}^{{r_{t}}}
\frac{{\rho}\left[\sigma_{r}^{2}\left({r}^{2}-R^{2}\right)+(\sigma_{t}^{2}/2)R^{2}\right]}
{{r}\sqrt{{r}^{2}-R^{2}}}d{r}
\end{equation}
where 
\begin{equation}\Sigma(R)=2\int_{R}^{{r_{t}}}\frac{{\rho}{r}}{\sqrt{{r}^{2}-R^{2}}}d{r}\end{equation}
is the projected density at $R$ and $r_t$ is the tidal radius.  The
above procedure allows to calculate a LOS velocity dispersion profile
for any given choice of $M/L$ and $\beta(r)$ according to both
Newtonian and MOND theories. 

We adopted the density profile of the King (1966) model that best-fits
the data of S11 defined by the parameters
$(W_{0},r_{c},\mu_{V,0})=(7,0.61\arcmin, 25.04~mag~arcsec^{-2})$ which
have been converted in physical units assuming a distance of $d=71$
kpc and two different values of the mass-to-light ratio: $M/L_{V}$=1.885
(derived for Pal 14 from stellar population synthesis by McLaughlin \&
van der Marel 2005) and $M/L_{V}$=0.747 (corresponding to the minimum mass
estimated by J09 from star counts in the color-magnitude
diagram). Hereafter, we will refer to the M/L ratio in the V band simply as M/L. 
Regarding the degree of anisotropy, we considered, besides
the isotropic case ($\beta=0$), two extreme anisotropic cases: purely
tangential ($\beta=-\infty$) and maximally radial.  In the latter case
we adopted the Osipkov-Merritt parametrization (Osipkov 1979; Merritt
1985)
\begin{equation}
\beta({r})=\frac{{r}^{2}}{{r}^{2}+{r}_{a}^{2}},
\end{equation}
where ${r}_{a}$ is the anisotropy radius, which sets the boundary
where orbits become significantly radially biased (systems with
smaller ${r}_{a}$ are more radially anisotropic).  We assumed
${r}_{a}={r}_{a,min}$, where ${r}_{a,min}$ is the bona-fide minimum
value for of ${r}_{a}$ for stability against radial-orbit
instability. In particular, we adopt ${r}_{a,min}=2.8 r_c$ for the
Newtonian model and ${r}_{a,min}=3.1 r_c$ for the MOND one,
corresponding to the marginal condition for stability
$2T_{r,half}/T_{t,half}\simeq 1.5$ (where $T_{r,half}$ and
$T_{t,half}$ are the radial and tangential kinetic energies computed
within the half-mass radius $r_{half}$; Nipoti et al. 2011; Ibata et
al. 2011a).

The LOS velocity dispersion profiles of all these models, calculated
by integrating numerically the above equations, are shown in
Fig. \ref{sig}.  It is evident that, at least as long as the cluster
is treated as isolated, MOND models predict a significantly larger
velocity dispersion with respect to Newtonian ones (see also Baumgardt
et al. 2005; Sollima \& Nipoti 2010; Haghi et al. 2009,
2011). Newtonian and MOND velocity dispersion profiles are clearly
distinct, regardless of the assumed anisotropy profile.

\subsection{N-body simulations}
\label{nbody_sec}

As already reported above, all the models presented in the previous
section are based on the assumption that the cluster is isolated. 
However, it is known that the presence of the Galactic field is important
for Pal 14 (see Sect. \ref{intro_sec}), with different effects
depending of the considered gravity law. In particular:
\begin{itemize}
\item{In Newtonian gravity, the tidal interaction with the Galactic
  potential can alter the velocity dispersion of the system by
    heating stars at large distances from the center, by producing virial
    oscillations after tidal shocks and influencing its structural
    evolution;}
\item{In MOND, besides the above tidal effects, even the presence of a
    uniform external field breaks the spherical symmetry of the
    gravitation field and can lead the cluster toward a less deep MOND
    regime (Bekenstein \& Milgrom 1984).}
\end{itemize} 
To account properly for the above effects we used a set of N-body
simulations performed in both Newtonian gravity and MOND.  In the
Newtonian case we studied the tidal effects by simulating the
evolution of the cluster orbiting within the Galactic potential.
  In the MOND case we used N-body simulations to model the cluster in
  the presence of a uniform external field, so for simplicity we
  neglected the tidal effects. We note that, given the very long
two-body relaxation time of Pal 14 ($t_{rh}\sim19.9\,Gyr$; S11), its
time evolution can be simulated also with collisionless N-body codes,
which is especially convenient in MOND, because of the difficulty in
realizing a collisional N-body code in this non-linear theory.

\subsubsection{Newtonian simulations}
\label{newt_sec}

\begin{table}
\begin{center}
\caption{Initial conditions of the Newtonian N-body simulations: orbital
eccentricity ($e$), orbital energy ($E$), obital angular momentum ($L_{z}$),
cluster mass (M), King (1966) central adimensional potential ($W_{0}$) and core
radius ($r_{c}$).}
\label{tab:init_cond}
\begin{tabular}{lccccccr}
\tableline\tableline
Model & $e$ & E           & $L_{z}$  & $P_{orb}$ & M           & $W_{0}$ & $r_{c}$\\
\#    &     & ($km/s^{2}$)  & (km/s kpc) & (Gyr)       & ($M_{\odot}$) &         & (pc)\\
\tableline
37/39 & 0.002 & 15962.05 & 9100 & 2.04 & 50000 & 9 & 12.6\\
41/43 & 0.5 & 5661.36  & 6300 & 2.23 & 380000 & 12 & 12.6\\
45/47 & 0.002 & 15962.05 & 9100 & 2.04 & 105000 & 12 & 12.6\\
49/51 & 0.5 & 5661.36  & 6300 & 2.23 & 170000 & 12 & 12.6\\
\tableline\tableline
\end{tabular}
%\tablecomments{}
\end{center}
\end{table}

\begin{figure*}
\epsscale{1.}
\plotone{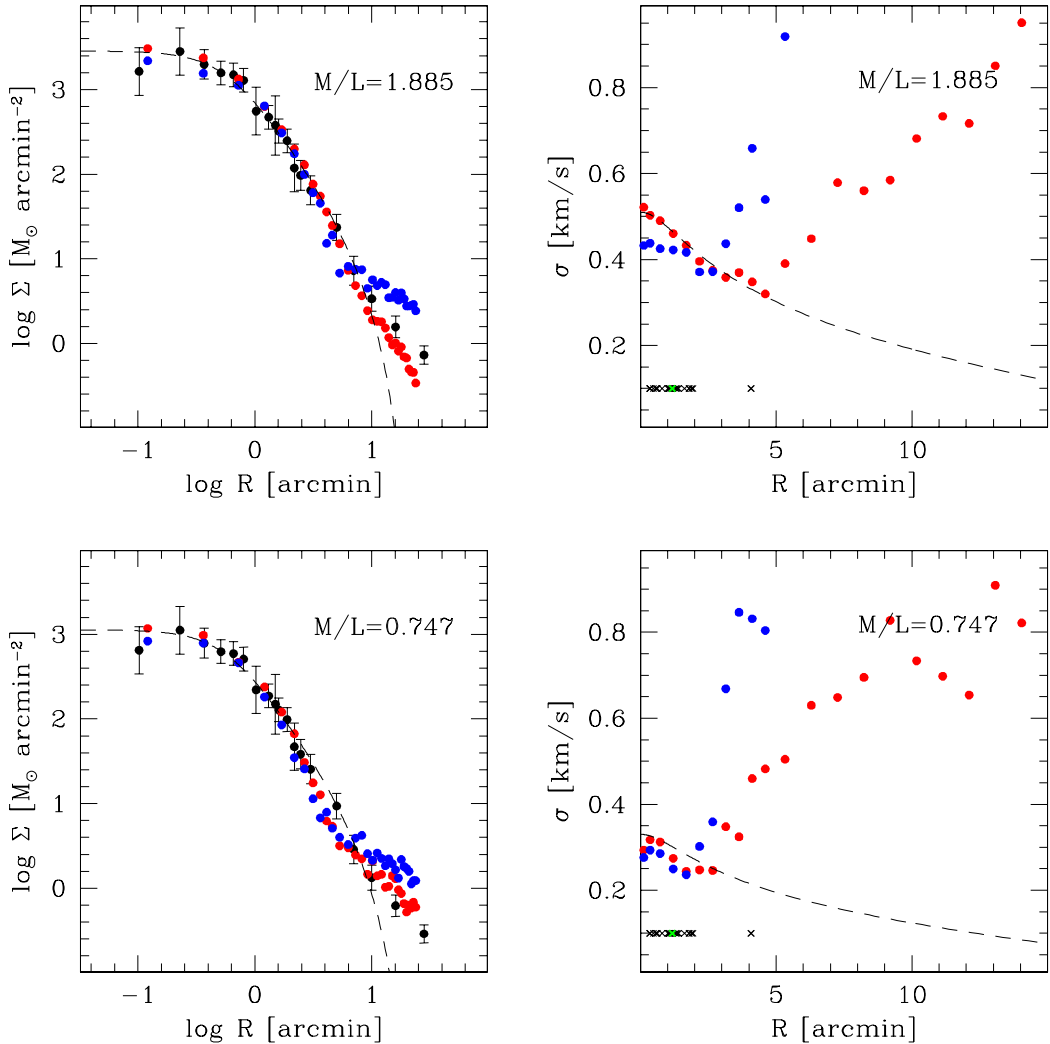}
\caption{Projected density profiles (left panels) and LOS velocity
  dispersion profiles (right panels) of the outcome of the N-body
  simulations performed in Newtonian dynamics assuming a
  quasi-circular orbit ($e=0.002$; red points) and an eccentric orbit
  ($e=0.5$; blue points). Upper panels refer to M/L=1.885, bottom
  panels to M/L=0.747. The density profile measures of S11 are marked
  with filled points with error bars in the left panels.  The radial
  locations of the target stars are marked at the bottom of the right
  panel with black crosses, the green cross indicates the location of
  star \#15. The predictions of the isolated Newtonian King (1966)
  model are also shown in all panels with dashed lines.}
\label{tides1}
\end{figure*}

As already discussed in Sect. \ref{nbody_sec}, the tidal interaction
between Pal 14 and the Milky Way is expected to heat the 
  outskirts of the cluster. This can be the result of both
compressive shocks occurred during the disk crossing and perigalactic
passages (Ostriker et al. 1972; Gnedin \& Ostriker 1997) and the
sudden change in the underlying potential (like in the case of the
Sagittarius galaxy; Taylor \& Babul 2001). After these episodes, the
kinetic energy of a fraction of the cluster stars can exceed the
boundaries of the cluster potential well and these stars become
"potential escapers" (Kupper et al. 2010). These stars might have a
velocity dispersion larger than the other bound stars and remain
within the cluster in spite of their positive energy for a timescale
comparable to the cluster orbital period (Lee \& Ostriker 1987).  
  Moreover, the strong perturbations occurring after every disk
  passage affect the cluster virial equilibrium (and consequently its
  internal kinematics) producing damped oscillations which go on for
  many dynamical times (Gnedin \& Ostriker 1999).  Finally, the
  structural evolution of the cluster is also influenced by tides
  which accelerate the process of mass-loss (Gnedin et al. 1999).  The
  overall effect on the cluster velocity dispersion is therefore
  extremely complex and not obviously resulting in a heating/freezing.

To evaluate the effect of tides on the velocity dispersion of Pal 14
we ran a set of N-body simulations. We used the last version of
NBSymple (Capuzzo-Dolcetta et al. 2011), an efficient N-body
integrator implemented on a hybrid CPU + GPU platform exploiting a
double-parallelization on CPUs and on the hosted Graphic Processing
Units (GPUs). The precision is guaranteed resorting to direct
summation (to avoid truncation errors in force evaluation), and on the
usage of high order, symplectic time integration methods (Kinoshita et
al. 1991; Yoshida 1991).  In particular the code allows to choose
between two different symplectic integrators: a second order algorithm
(commonly known as leapfrog) and a much more accurate (but also time
consuming) sixth order method. The effect of the external galactic field is taken into
account using an analytical representation of its gravitational
potential.  We adopted a leap-frog scheme with a time-step of $\Delta
t=3.7\times 10^{4}~yr$ and a softening length of 0.2 pc (following the
prescription of Dehnen \& Read 2011). Such a relatively large
time-step and softening length do not affect the accuracy of the
simulation, because, as mentioned above, the effects of two-body
encounters has been found to be negligible in this cluster even in its
innermost region (S11; Beccari et al. 2011) and the relaxation times
at the half-mass radius of our models are larger than the cluster age
during the entire simulation. The cluster was
launched within the three-component (bulge + disc + halo) static
Galactic potential of Johnston et al. (1995) on two orbits with
different eccentricities ($e\sim 0$ and $e=0.5$). First of all, we
integrated backward in time, within the adopted potential, the orbit
of a test particle from the apocenter (assumed to be the present-day
location of the cluster at R=45.75 kpc, z=47.68 kpc, in cylindrical
coordinates centered on the Galactic center) to the epoch $t=
2~P_{orb}$ ago\footnote{We conducted the simulations only for the last
  two orbital periods as potential escaper stars generated in previous
  orbits are expected to be already evaporated from the cluster, thus
  not affecting its kinematics.}. At the end of the simulation, the
model is within 40 pc of the current position of Pal 14 and the energy
has been conserved within $\Delta E/E\sim 10^{-7}$ for both the
considered orbits.  Then the simulation has been launched with a
cluster represented by 61,440 equal-mass particles\footnote{The
  adoption of equal-mass particles instead of a mass spectrum is not
  expected to affect the structural and dynamical evolution of the
  cluster because of the lack of relaxation in Pal 14 (S11; Beccari et
  al. 2011).} distributed according to an equilibrium King (1966)
model having initial mass, core radius and concentration empirically
determined so that the projected density profile after two complete
orbits reproduces the present-day cluster configuration (see
Sect. \ref{analyt_sec}). The initial conditions of the simulation for
the two adopted orbits are reported in Table 1.

In Fig. \ref{tides1} the final projected density and LOS velocity
dispersion profiles are shown for the considered combinations of
  values of orbital eccentricity and M/L. It can be noted that 
  in all cases the effect of tidal heating is visible only in the
outermost radial bins of the projected density profiles, but a
significant deviation from the predicted behavior of the isolated
model is clearly visible in the velocity dispersion profiles beyond a
distance from the cluster center which depends on the orbital
eccentricity and the adopted M/L ratio. On the other hand, the isolated model remains a good
representation of the cluster velocity dispersion at small distance
from the cluster center, where most of the stars of the J09 sample
reside.

\subsubsection{MOND simulations}
\label{mond_sec}

\begin{figure}
\epsscale{1.2}
\plotone{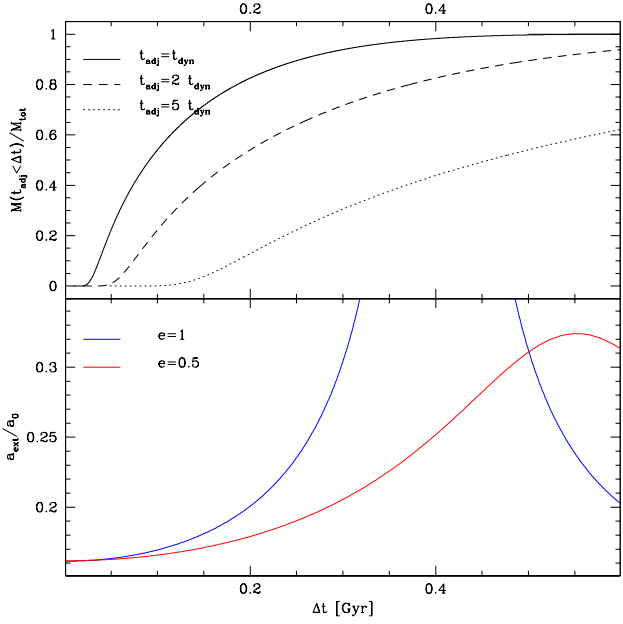}
\caption{Bottom panel: External acceleration exerted by the Milky Way
  on Pal 14 as a function of time elapsed from the present-day
  position (which is assumed to be the apocenter) for two different
  orbital eccentricities (e=1: grey line; e=0: black line; colored
  blue and red in the online version). Top panel: fraction of cluster
  stars with a timescale of adjustment with the external field
  ($t_{adj}$) shorter than the time elapsed from the initial
  position. The cases of $t_{adj}/t_{dyn}$=1, 2 and 5 are shown with
  solid, dashed and dotted lines, respectively. For instance in the
  case e=0.5, $\sim 0.56$ Gyr before the apocentric passage the
  modulus of the external field is twice the apocentric value (lower
  panel), but this time interval corresponds to more than 2 dynamical
  times for $\sim 90\%$ of the stars of the clusters (upper panel),
  which are thus likely to be now already in equilibrium with the
  current external field.}
\label{orb}
\end{figure}

\begin{figure*}
\epsscale{1.}
\plotone{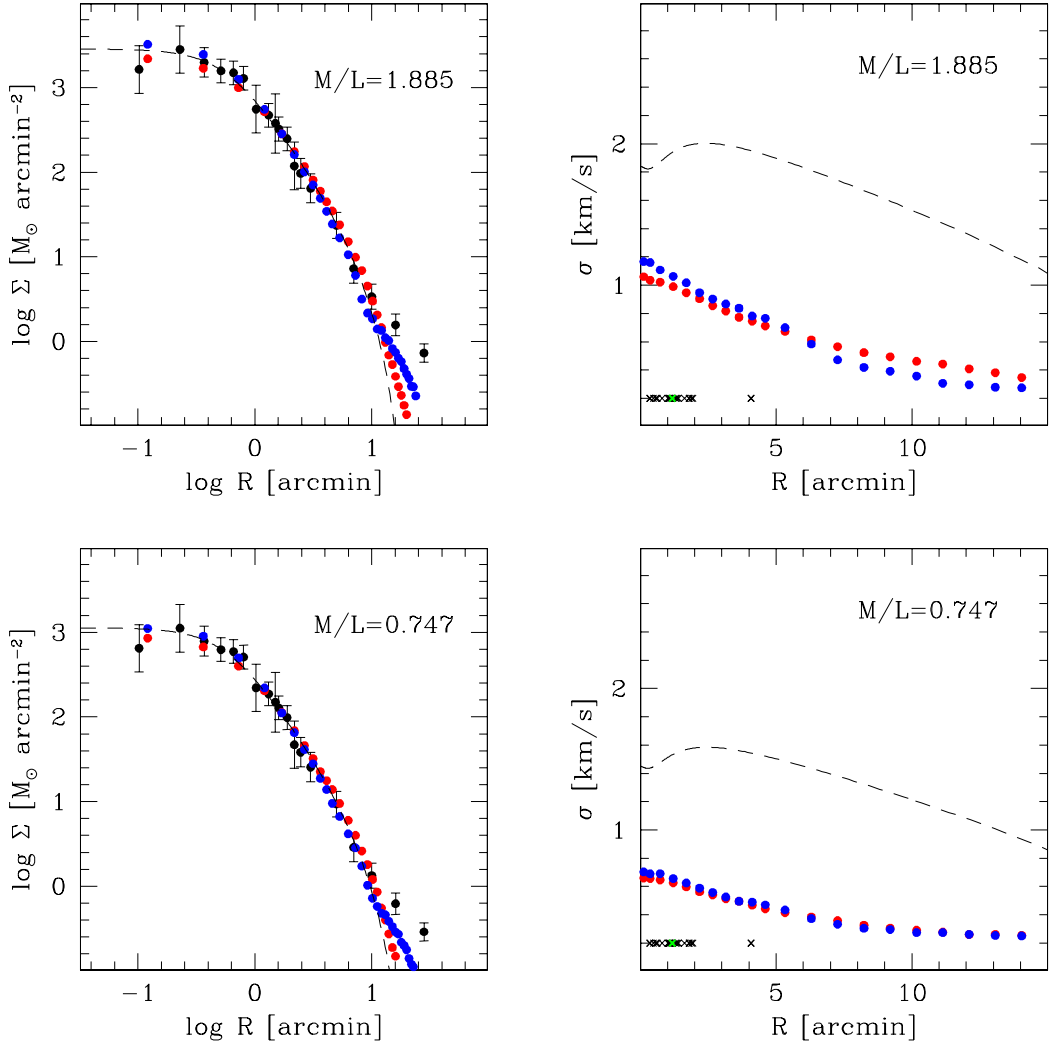}
\caption{Projected density profiles (left panels) and LOS velocity
  dispersion profiles (right panels) of the outcome of the $e=0$ (red
  points) and $e=0.5$ (blue points) MOND N-body simulations (see text
  for details). Upper panels refers to M/L=1.885, bottom panels to
  M/L=0.747. The density profile measures of S11 are marked with
  filled points with error bars in the left panel. The radial
  locations of the target stars are marked at the bottom of the right
  panel with black crosses, the green cross indicates the location of
  star \#15.  The corresponding isolated MOND models are represented
  in all panels with dashed lines.}
\label{efe1}
\end{figure*}

The gravitational field of the Galaxy at the location of Pal 14 is
estimated to be, in modulus, $g_{ext}\sim0.16 a_0$ (Johnston et
al. 1995) and directed approximately towards the Galactic center.
During its orbit the cluster will occupy different Galactic regions
where the external field differs both in direction and in magnitude.
The effect of such a variation on the internal kinematics of the
cluster can be in principle significant (Kosowsky 2010) and is linked
to the ratio between the timescale necessary to the cluster to reach 
a new equilibrium configuration ($t_{adj}$; which is of the order of 
the internal dynamical time $t_{dyn}$) and the
orbital period ($P_{orb}$).  Although the orbit of Pal 14 is unknown
we can set an upper limit to this ratio by assuming the cluster
presently at its apocenter and considering different orbital
eccentricities.  As a first order approximation, we calculated the
orbital period assuming planar orbits within a spherical isothermal
Galactic potential $\Phi_{MW}=v_{circ}^{2} \ln d_{GC}$ (where $d_{GC}$
is the Galactocentric distance) with $v_{circ}=198.9~km/s$, so
$g_{ext}=0.16~a_{0}$ at $d_{GC}=66~ kpc$ (the present-day
Galactocentric distance of Pal 14).  The dynamical time has been
calculated as
$$t_{dyn}(r)=2 \pi\sqrt{\frac{r}{d\Phi(r)/d r}}$$ adopting the cluster
potential $\Phi(r)$ as in the isotropic MOND model of Pal 14 with
$M/L$=1.885.  

The ratio $t_{dyn}/P_{orb}$ is relatively small: even in the extreme
case of a free falling orbit ($e=1$) we get $t_{dyn}/P_{orb}=0.23$ at
the half-mass radius.  To visualize the effect of the variation of the
external acceleration on the internal cluster kinematics, we plot in
Fig. \ref{orb} the external acceleration exerted on the cluster after
a time interval $\Delta t$ from the current position (bottom panel)
for two different eccentric orbits ($e=0.5$ and 1) and the fraction of
mass comprised within the radius where $t_{adj}/t_{dyn}(r)=$1, 2 and
5.  It is evident that assuming $t_{adj}/t_{dyn}$=1, more than 80\% of
the cluster stars reach the equilibrium with the external field within
100 Myr, a timescale during which the cluster feels an almost constant
external acceleration (always smaller than $0.2~a_{0}$), regardless
of its orbit.  A more significant variation is instead noticeable
assuming $t_{adj}$ of the order of few dynamical times: if
$t_{adj}/t_{dyn}$=5 the external field doubles in the case of an orbit
with eccentricity $e=0.5$ before $\sim40\%$ of the cluster stars
reached equilibrium, and the effect can be even more significant in
the case of the free-falling orbit.

To evaluate the effect of the external field on the predicted velocity
dispersion we used the MOND N-body code {\sc n-mody} (Londrillo \&
Nipoti 2009), in which we implemented the external field effect as
described in Ibata et al. (2011a).  The presence of an uniform
external field $-\nabla \Phi_{ext}$ is accounted for in the boundary
conditions of equation~(\ref{mond_eq}), which become $\nabla \Phi \to
\nabla \Phi_{ext}$ for $|{\bf r}|\to \infty$ (Bekenstein \& Milgrom
1984).  We recall that, due to technical difficulties, in MOND we do
not simulate self-consistently the evolution of Pal 14 orbiting in the
Galaxy (including tidal effects), but we limit ourselves to
  simple numerical experiments that should allow us to estimate
  reasonably well the effect of a uniform external field on the
  cluster velocity dispersion.  In the simplest of our experiments we
simulate the cluster in the presence of a uniform external field ${\bf
  g}_{ext}=-\nabla \Phi_{ext}$ that remains constant in modulus and
direction throughout the simulation (see also Ibata et al. 2011a). In
particular, in the attempt to model the cluster on a circular orbit,
we fix $g_{ext}=0.16a_0$ (the value of the external field at the
present-day location of Pal~14). In these simulations (which hereafter
we refer to, loosely speaking, as $e=0$ MOND simulations) we let the
cluster evolve for $\sim 2.3\,Gyr$ (i.e. slightly more than a complete
orbital period).

If the cluster is on an eccentric orbit, not only the direction, but
also the modulus of the external field varies during the orbit. In
particular, if Pal 14 is currently at the apocenter it must have
experienced a stronger external field in the past, which might affect
to some degree its present-day velocity dispersion.  In order to
explore this effect, we also ran simulations in which the external
field varies in time (in modulus, but not in direction) as expected
for the $e=0.5$ orbit discussed above (see Fig.~\ref{orb}, bottom
panel). In these simulations we follow the evolution of the cluster
for $\sim 1.12\,Gyr$ (i.e. half orbit, starting at the apocenter): so,
in this case the external field is $g_{ext}\simeq 0.16 a_0$ at $t=0$ ,
it reaches a its maximum $g_{ext}\simeq 0.32 a_0$ at $\sim 0.56\,Gyr$
and at the end of the simulation is again $g_{ext}\simeq 0.16 a_0$
(loosely speaking hereafter we refer to these simulations as the
$e=0.5$ MOND simulations).

For each value of $M/L$ ($M/L=0.747$ and $M/L=1.885$) we ran both the
$e=0$ and the $e=0.5$ MOND simulations. In all cases, we initialized
our N-body simulations with quasi-equilibrium distributions of
$8\times10^5$ particles: in practice we generate the initial conditions as
follows. We first build an equilibrium Newtonian
model with density given by the best-fitting King model of Pal 14
described in Section \ref{mod_sec}, and we then multiply the velocity
of each particle by a factor
$\left[\left(1+\tilde{g}_{ref}\right)/\tilde{g}_{ref}\right]^{1/2}$
(trying to reproduce the expected ``quasi-Newtonian'' behavior), where
$\tilde{g}_{ref}\equiv g_{ref}/a_0$, and $g_{ref}$ is a reference
gravitational field modulus.
The value of $g_{ref}$ is empirically chosen in each case in
order to have at the end of the simulation a mock cluster with a 
surface-density profile consistent
with the observed profile of Pal 14.  In the case of the $e=0$
simulations $g_{ref}=g_{ext}=0.16 a_0$; in the case of the $e=0.5$
simulations $g_{ref}=\langle{g_{ext}}\rangle=0.22 a_0$, where
$\langle{g_{ext}}\rangle$ is the time-averaged modulus of the external
field for the considered orbit.

In the $e=0$ simulations the system readjusts itself in a few
dynamical times into a new equilibrium configuration.  In the $e=0.5$
simulations the cluster is
continuously slowly evolving due to the
variation of the external field. In both cases the end-products of the
simulations are axisymmetric configurations with symmetry axis along
the direction of the external field. When projected along the
line-of-sight to Pal 14 (assuming that the field of the Galaxy points
toward the Galactic center) the systems appear almost circular (with
ellipticity $\epsilon\sim 0.01$) and with circularized surface-density
distribution very similar to the observed light distribution of Pal 14
(see Fig.~\ref{efe1}, left panels).  In Fig.~\ref{efe1} we also show
the velocity dispersion profiles of the end-products of the $e=0$ and
$e=0.5$ MOND simulations (right panels).  It is apparent that when the
external field is taken into account the velocity dispersion profile
is at all radii significantly lower than that of the corresponding
isolated model. As we will see in Sect. \ref{resmond_sec}, this effect
is crucial in the comparison between the MOND predictions and the
data, in particular when low values of $M/L$ are assumed.  On the
other hand, the orbital eccentricity does not affect significantly the
velocity dispersion profile, indicating that the cluster quickly
reaches the equilibrium with the external field at the apocenter and
does not keep memory of the stronger external field experienced in the
past.

\subsection{Binary population}
\label{bin_sec}

It is well known that a significant population of undetected binaries
can inflate the measured velocity dispersion (Blecha et al. 2004),
because in a binary system the relative projected velocity of the
primary component is added to the motion of the center of mass,
introducing an additional spread in the velocity distribution of the
whole population.  To account for the effect of a binary population on
the velocity dispersion, we constructed a library of binaries which
has been used in the Monte Carlo procedure.

Following McConnachie \& Cote (2010) the projected velocity of the
primary component in a binary system is given by
$$v=\frac{2\pi a_{1} \sin i}{P(1-e^{2})^{1/2}}[\cos(\theta+\omega)+e \cos
\omega],$$ 
where $m_{1}$ and $m_{2}$ are the masses of the primary and
secondary component, $a_{1}$ is the semi-major axis of the primary
component, $P$ is the orbital period, $e$ the eccentricity, $i$ the
inclination angle to the line-of-sight, $\theta$ the phase from the
periastron and $\omega$ the longitude of the periastron. For each
simulated binary we extracted randomly a combination of
($m_{1},m_{2}, P,e,\theta,\omega,i$) from suitable distributions and
derived the corresponding projected velocity $v$.

We randomly extracted a large number ($N>10^{6}$) of stars from the
IMF of Kroupa (2002) with masses
$0.08<M/M_{\odot}<7$ and paired them imposing that the distribution of
mass ratios in the primary component mass range $1<m_{1}/M_{\odot}<7$
must be equal to that measured by Fisher et al. (2005) in the solar
neighborhood. To do this, an iterative algorithm has been used: at
every iteration stars are paired randomly and a "chance of pairing" as
a function of the mass ratio has been computed as the ratio between
the output mass ratio normalized distribution calculated in the above
primary component mass range and that of Fisher et
al. (2005)\footnote{Of course, in this algorithm we implicitly assume
  that the chance of pairing depends only on the mass ratio. Although
  this assumption is clearly a simplification, it allows to reproduce
  the observed mass-ratio distribution in star clusters better than a
  simple random pairing (Sollima et al. 2010).}.  Then, a subsample of
binaries has been extracted according to the chance of pairing
associated to their mass ratio and added to the library, while the
components of the rejected binaries are used as input in the next
iteration until all stars are paired. Then, we selected from the
library all binaries whose primary component is within
$|m_{1}-M_{RGB}|<0.05 M_{\odot}$, where $M_{RGB}=0.83~M_{\odot}$ is
the typical mass of a RGB star calculated by comparing the
color-magnitude diagram of Pal 14 of S11 with a suitable isochrone of
Marigo et al.  (2008).  We followed the prescriptions of Duquennoy \&
Mayor (1991) for the distribution of periods and eccentricities. The
semi-major axis has been calculated using the third Kepler law
$$a_{1}=\frac{1}{1+\frac{m_{1}}{m_{2}}}\left[\frac{P^{2}G(m_{1}+m_{2})}{4\pi^{2}}\right]^{1/3}.$$
We removed all those binaries whose corresponding semi-major axes lie
outside the range $a_{min}<a<100$~AU where $a_{min}$ is linked to the
radius of the secondary component (according to Lee \& Nelson
1988). The distribution of the angles ($i,\theta,\omega$) has been
chosen according to the corresponding probability distributions (${\rm
  Prob}(i)\propto \sin~i;~{\rm Prob}(\theta)\propto
\dot{\theta}^{-1};~ {\rm Prob}(\omega)={\rm constant}$).

For a given fraction of binaries $f_{b}$ the effective fraction 
$f_{b}^{RGB}$ is calculated as the ratio between the number of binaries with 
$|m_{1}-M_{RGB}|<0.05 M_{\odot}$ ($N_{b,RGB}$) and the number of objects (singles+binaries)
in the same mass range ($N_{s,RGB}+N_{b,RGB}$) as
$$f_{b}^{RGB}=\frac{f_{b} N_{b,RGB}}{(1-f_{b})N_{s,RGB}\frac{N_{b,tot}}{N_{s,tot}}+f_{b}
N_{b,RGB}}.$$

Note that in our Monte Carlo simulations we assumed that the binary
population has the same radial distribution as single stars. This
assumption is justified by the observed lack of mass segregation
observed in this cluster (Beccari et al. 2011).

\section{Monte Carlo simulations}

\begin{table*}
\begin{center}
\caption{Summary of the Monte Carlo simulations. The table lists the model ID, the adopted theory of gravity, the kind of
anisotropy, the M/L ratio, the cluster orbital eccentricity ($e$), the binary
fraction ($f_{b}$), the rejection threshold and 
the derived velocity dispersion quantiles 50\% (median), 16\% and 84\% ($\sim~1\sigma$)}
\label{tab:summ}
\begin{tabular}{lccccccccc}
\tableline\tableline
Model & Gravity & anisotropy & M/L                     & $e$ & $f_{b}$ & rejection threshold & \multicolumn{3}{c}{$\sigma_{v}$}\\
  \#  &         &            & $M_{\odot}/L_{V,\odot}$ &     &   \%  & $\sigma$           & f(16\%) & f(50\%) & f(84\%)\\
\tableline						    
1  & Newtonian &  isotropic        &  1.885  &   isolated   &  0 & 3  & 0.30 & 0.42 & 0.56\\	 
2  & MOND      &  isotropic        &  1.885  &   isolated   &  0 & 3  & 1.52 & 1.84 & 2.27\\	 
3  & Newtonian &  isotropic        &  1.885  &   isolated   &  0 & 5  & 0.32 & 0.41 & 0.58\\	 
4  & MOND      &  isotropic        &  1.885  &   isolated   &  0 & 5  & 1.45 & 1.90 & 2.19\\	 
5  & Newtonian &  isotropic        &  0.747  &   isolated   &  0 & 3  & 0.16 & 0.27 & 0.41\\	 
6  & MOND      &  isotropic        &  0.747  &   isolated   &  0 & 3  & 1.25 & 1.48 & 1.90\\	 
7  & Newtonian &  isotropic        &  0.747  &   isolated   &  0 & 5  & 0.15 & 0.29 & 0.40\\	 
8  & MOND      &  isotropic        &  0.747  &   isolated   &  0 & 5  & 1.19 & 1.50 & 1.80\\	 
9  & Newtonian &  tangential       &  1.885  &   isolated   &  0 & 3  & 0.38 & 0.48 & 0.67\\	 
10 & MOND      &  tangential       &  1.885  &   isolated   &  0 & 3  & 1.59 & 2.06 & 2.33\\	 
11 & Newtonian &  tangential       &  1.885  &   isolated   &  0 & 5  & 0.39 & 0.49 & 0.68\\	 
12 & MOND      &  tangential       &  1.885  &   isolated   &  0 & 5  & 1.72 & 1.97 & 2.47\\	 
13 & Newtonian &  radial           &  1.885  &   isolated   &  0 & 3  & 0.31 & 0.47 & 0.59\\	 
14 & MOND      &  radial           &  1.885  &   isolated   &  0 & 3  & 1.60 & 2.04 & 2.41\\	 
15 & Newtonian &  radial           &  1.885  &   isolated   &  0 & 5  & 0.32 & 0.45 & 0.59\\	 
16 & MOND      &  radial           &  1.885  &   isolated   &  0 & 5  & 1.56 & 1.95 & 2.37\\	 
17 & Newtonian &  isotropic        &  1.885  &   isolated   & 10 & 3  & 0.26 & 0.49 & 0.68\\	 
18 & Newtonian &  isotropic	   &  1.885  &   isolated   & 10 & 5  & 0.32 & 0.51 & 0.91\\ 
19 & Newtonian &  isotropic        &  0.747  &   isolated   & 10 & 3  & 0.13 & 0.31 & 0.50\\	 
20 & Newtonian &  isotropic	   &  0.747  &   isolated   & 10 & 5  & 0.16 & 0.32 & 0.77\\ 
21 & Newtonian &  isotropic	   &  1.885  &   isolated   & 20 & 3  & 0.27 & 0.57 & 0.94\\
22 & Newtonian &  isotropic	   &  1.885  &   isolated   & 20 & 5  & 0.42 & 0.77 & 1.20\\	 
23 & Newtonian &  isotropic	   &  0.747  &   isolated   & 20 & 3  & 0.01 & 0.34 & 0.61\\
24 & Newtonian &  isotropic	   &  0.747  &   isolated   & 20 & 5  & 0.24 & 0.59 & 1.09\\	 
25 & Newtonian &  isotropic	   &  1.885  &   isolated   & 30 & 3  & 0.36 & 0.71 & 1.33\\	 
26 & Newtonian &  isotropic	   &  1.885  &   isolated   & 30 & 5  & 0.65 & 1.00 & 1.47\\	 
27 & Newtonian &  isotropic	   &  0.747  &   isolated   & 30 & 3  & 0.15 & 0.56 & 1.12\\	 
28 & Newtonian &  isotropic	   &  0.747  &   isolated   & 30 & 5  & 0.48 & 0.92 & 1.39\\	 
29 & Newtonian &  isotropic	   &  1.885  &   isolated   & 40 & 3  & 0.61 & 1.15 & 1.70\\ 
30 & Newtonian &  isotropic	   &  1.885  &   isolated   & 40 & 5  & 0.84 & 1.17 & 1.74\\		 
31 & Newtonian &  isotropic	   &  0.747  &   isolated   & 40 & 3  & 0.31 & 0.66 & 1.41\\ 
32 & Newtonian &  isotropic	   &  0.747  &   isolated   & 40 & 5  & 0.74 & 1.14 & 1.71\\		 
33 & Newtonian &  isotropic        &  1.885  &   isolated   & 50 & 3  & 0.98 & 1.43 & 2.05\\		  
34 & Newtonian &  isotropic        &  1.885  &   isolated   & 50 & 5  & 1.04 & 1.52 & 1.93\\	 
35 & Newtonian &  isotropic        &  0.747  &   isolated   & 50 & 3  & 0.59 & 1.31 & 1.77\\		  
%
%\tableline\tableline
%\end{tabular}
%%\tablecomments{}
%\end{center}
%\end{table*}
%
%\addtocounter{table}{-1}
%\begin{table*}
%\begin{center}
%\caption{(continued)}
%%\label{tab:summ}
%\begin{tabular}{lccccccccc}
%\tableline\tableline
%Model & Gravity & anisotropy & M/L                     & $e$ & $f_{b}$ & rejection threshold & \multicolumn{3}{c}{$\sigma_{v}$}\\
%  \#  &         &            & $M_{\odot}/L_{V,\odot}$ &     &   \%  & $\sigma$           & f(16\%) & f(50\%) & f(84\%)\\
%\tableline						    
36 & Newtonian &  isotropic        &  0.747  &   isolated   & 50 & 5  & 0.77 & 1.31 & 1.69\\	 
37 & Newtonian &  isotropic        &  1.885  &   0.002      &  0 & 3  & 0.31 & 0.42 & 0.59\\
38 & MOND      &  isotropic        &  1.885  &   0	    &  0 & 3  & 0.80 & 0.95 & 1.23\\
39 & Newtonian &  isotropic        &  1.885  &   0.002      &  0 & 5  & 0.32 & 0.46 & 0.58\\
40 & MOND      &  isotropic        &  1.885  &   0	    &  0 & 5  & 0.76 & 0.97 & 1.20\\ 
41 & Newtonian &  isotropic        &  1.885  &   0.5        &  0 & 3  & 0.30 & 0.42 & 0.57\\
42 & MOND      &  isotropic        &  1.885  &   0.5        &  0 & 3  & 0.61 & 0.81 & 1.14\\
43 & Newtonian &  isotropic        &  1.885  &   0.5        &  0 & 5  & 0.29 & 0.39 & 0.57\\
44 & MOND      &  isotropic        &  1.885  &   0.5        &  0 & 5  & 0.60 & 0.83 & 1.13\\
45 & Newtonian &  isotropic        &  0.747  &   0.002      &  0 & 3  & 0.14 & 0.27 & 0.37\\
46 & MOND      &  isotropic        &  0.747  &   0	    &  0 & 3  & 0.45 & 0.59 & 0.76\\
47 & Newtonian &  isotropic        &  0.747  &   0.002      &  0 & 5  & 0.15 & 0.28 & 0.38\\
48 & MOND      &  isotropic        &  0.747  &   0	    &  0 & 5  & 0.48 & 0.59 & 0.78\\
49 & Newtonian &  isotropic        &  0.747  &   0.5        &  0 & 3  & 0.11 & 0.25 & 0.41\\
50 & MOND      &  isotropic        &  0.747  &   0.5        &  0 & 3  & 0.32 & 0.54 & 0.71\\
51 & Newtonian &  isotropic        &  0.747  &   0.5        &  0 & 5  & 0.01 & 0.29 & 0.36\\
52 & MOND      &  isotropic        &  0.747  &   0.5        &  0 & 5  & 0.33 & 0.51 & 0.72\\

\tableline\tableline
\end{tabular}
%\tablecomments{}
\end{center}
\end{table*}

\begin{figure}
\epsscale{1.2}
\plotone{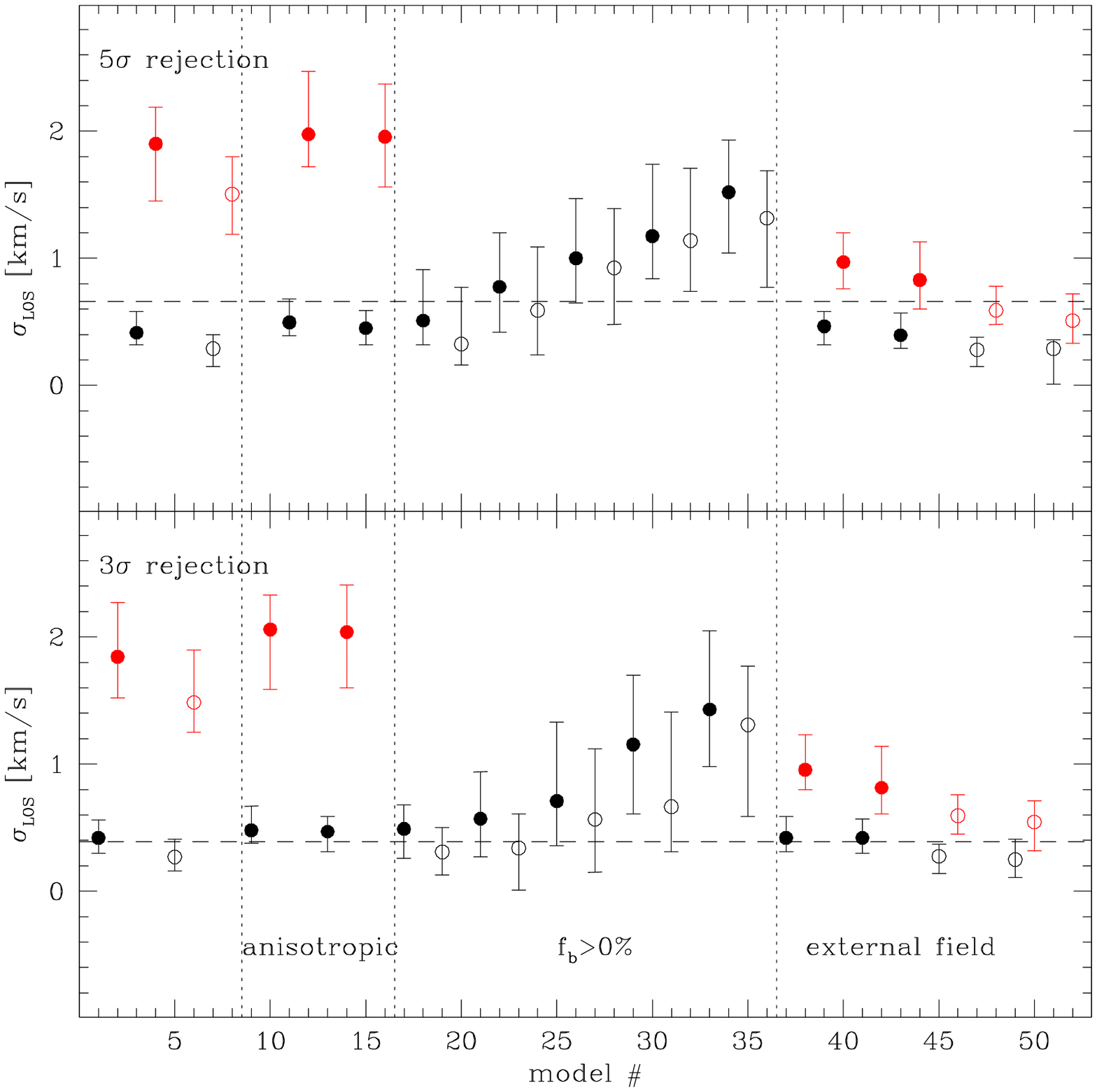}
\caption{Projected velocity dispersion and range comprising 68\% of 
realizations predicted by the whole set of Monte Carlo simulations. 
Black dots mark the prediction of Newtonian models, grey dots 
(red in the online version) the predictions of MOND models. 
Filled symbols refer to M/L=1.885, open symbols refer to M/L=0.747.}
\label{riass}
\end{figure}

To compare the velocity dispersion measured in Sect.~\ref{data_sec}
with the models described in Sect. \ref{mod_sec} we adopted a Monte
Carlo approach.  In particular, we first defined a rejection criterion
and selected a subsample of member stars from the J09 radial
velocities accordingly. Here, we adopted a preliminary exclusion of
all stars with $|v-\overline{v}|>5~km/s$, and a subsequent $\sigma$
clipping rejection. We considered the cases of a 3$\sigma$ and a
5$\sigma$ threshold---where $\sigma\equiv\sqrt{\sigma_{v}^{2}+
  \delta_{i}^{2}+\delta(\overline{v})^{2}}$ includes the intrinsic
velocity dispersion $\sigma_{v}$, the individual error $\delta_{i}$
and the error on the mean $\delta(\overline{v})$---which define two
samples with $N=16$ and $N=17$ stars (including star \#15),
respectively.  Then, once selected a model and a binary fraction
$f_{b}$, for each star of the observed sample the following steps have been
performed:
\begin{enumerate}
\item{We compute the distance of the star from the cluster center $R_{i}$ and
  extract a velocity corresponding to the local LOS velocity
  dispersion in the model. The extraction algorithm depends on the
  kind of model:
\begin{itemize}
\item{for the analytical models, the velocity is randomly extracted from
  a Gaussian distribution with dispersion $\sigma_{LOS}(R_{i})$ i.e. the
  projected velocity dispersion at the star distance from the center;}
\item{for the N-body models, the velocity is extracted from the last
  snapshot of the N-body simulation: for each star a random position
  angle is extracted and the projected velocity of the closest
  particle in the N-body system is adopted.}
\end{itemize}}
\item{We extract a random velocity from a Gaussian distribution with
  dispersion $\delta_{i}$ and sum it to the previous one to compute a
  simulated observational velocity;}
\item{A random number between 0 and 1 is extracted from a uniform
  distribution. If such number is smaller than the effective binary
  fraction $f_{b}^{RGB}$ (see Sect. \ref{bin_sec}) a binary is randomly
  extracted from the library defined in Sect. \ref{bin_sec} and its
  apparent velocity is added to the simulated velocity.}
\end{enumerate}
Then the rejection criterion is applied to the simulated sample and
for every rejected velocity another extraction is made at the same
location. In this way, every realization is constituted by the same
number of object at the same location as the observed sample.
Finally, the velocity dispersion of the simulated sample is calculated
using eq. \ref{sig_eq}.  The above procedure is repeated 1000 times to
compute the distribution of predicted velocity dispersions.
The outcome of all the performed simulations is summarized in Table 2 and
illustrated in Fig. \ref{riass}.

\newpage
\section{Results}

\subsection{Newtonian vs. MOND models}
\label{resmond_sec}

\begin{figure}
\epsscale{1.2}
\plotone{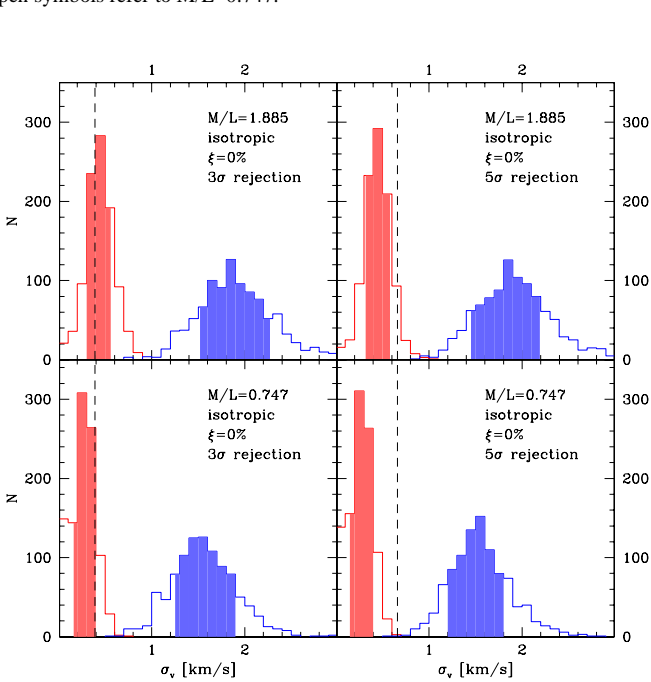}
\caption{Distribution of predicted velocity dispersions for Newtonian
  (red histograms) and MOND (blue histograms) isolated isotropic
  models with $f_{b}=0$ and $M/L$=1.885 (upper panels) and $M/L$=0.747
  (bottom panels). Left panels refer to the 3$\sigma$ rejection
  threshold, right panels to the 5$\sigma$ rejection threshold. Shaded
  areas indicate the range (centered on the median) comprising 68\% of
  realizations. The observed velocity dispersions are also marked in
  all panels with dashed lines.}
\label{ml}
\end{figure}

\begin{figure}
\epsscale{1.2}
\plotone{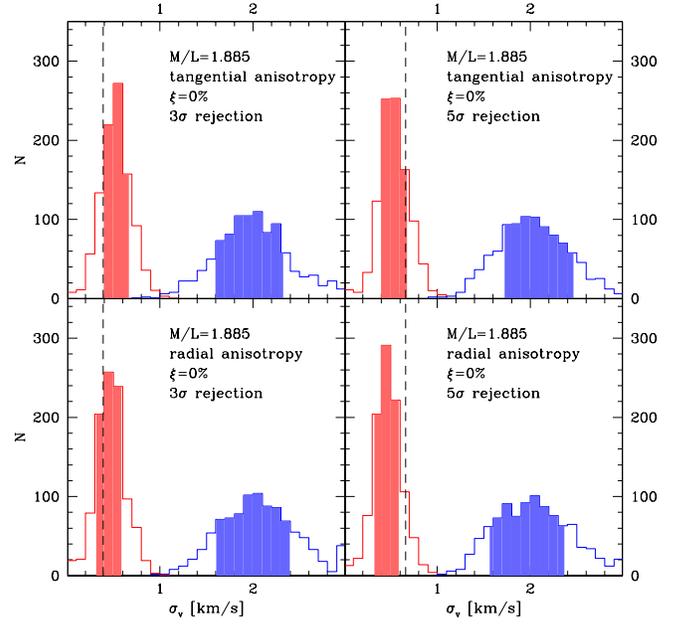}
\caption{Same as Fig. \ref{ml}, but for models purely tangential
  (upper panels) and extremely radial (bottom panels) anisotropy
  (assuming $M/L$=1.885).}
\label{anis}
\end{figure}

\begin{figure}
\epsscale{1.2}
\plotone{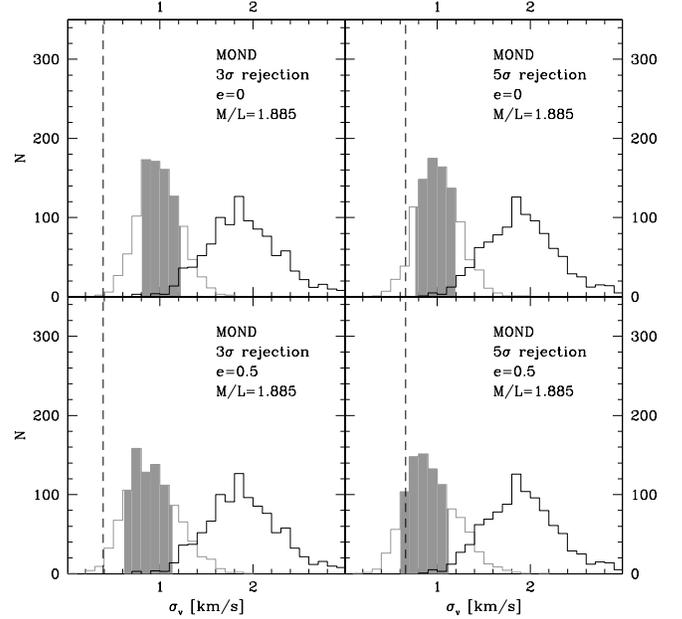}
\caption{Distribution of predicted velocity dispersions for isolated
  isotropic MOND models with $M/L$=1.885 and $f_{b}=0$ (black
  histograms) and models immersed in the external Galactic field (grey
  histograms), assuming
  a circular orbit (upper panels) and an eccentric orbit with e=0.5 (lower
  panels).  Left panels refer to the 3$\sigma$ rejection threshold,
  right panels to the 5$\sigma$ rejection threshold. Shaded areas
  indicate the range (centered on the median) comprising 68\% of
  realizations. The observed velocity dispersions are marked in all
  panels with dashed lines.}
\label{efe2}
\end{figure}

\begin{figure}
\epsscale{1.2}
\plotone{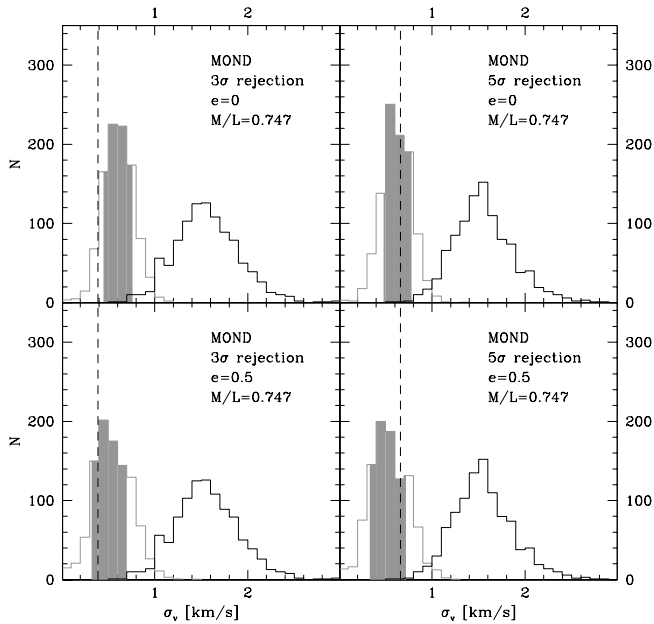}
\caption{Same as Fig. \ref{efe2}, but for models with $M/L$=0.747.}
\label{efe4}
\end{figure}

In Figs. \ref{ml} and \ref{anis} the distributions of predicted
velocity dispersions are shown for isolated models with no binaries
and different values of $M/L$ and degrees of anisotropy, respectively.
It is apparent that while Newtonian models with M/L=1.885 predict
velocity dispersions which are compatible with the observed value,
Newtonian models with a M/L=0.747 tend to underpredict the systemic
velocity dispersion, while all extractions from isolated MOND models
predict velocity dispersions larger than the observed (even when $M/L$
is as low as 0.747)\footnote{In Newtonian simulations with small
    M/L the intrinsic velocity dispersion is smaller the statistic
    uncertainty: as a consequence, in these cases a significant number
    of Monte Carlo realizations have $\sigma_{v}\sim 0$ (see
    Figs. \ref{ml} and \ref{tides3}).}. It is interesting to note
that anisotropy has only a small effect on the predicted distribution
of velocity dispersions. This is because most of the J09 sample is
constituted by stars located between $1\arcmin<R<2\arcmin$, a region
where all the models predict a similar velocity dispersion regardless
of the degree of anisotropy (see Fig. \ref{sig}).

In Fig. \ref{efe2} and \ref{efe4} the distributions of velocity
dispersions predicted by MOND models including the external field
effect (for the considered orbits and M/L ratios) are compared with
that of the isolated MOND model. As anticipated in
  Sect. \ref{mond_sec}, it is apparent that the inclusion of the
  external field significantly affects the predicted velocity
  dispersion reducing the systemic velocity dispersion.  For a
$M/L$=1.885 (the one predicted by stellar evolution models for Pal
14), it is clear that the predicted velocity dispersion, though
significantly influenced by the external field, remains in any case
higher than the observed value (Fig. \ref{efe2}). A small, but not
negligible ($\sim20$\%), number of realizations compatible with
observations are noticeable when a stronger external field and a
5$\sigma$ rejection threshold are considered. Note that a non-zero
binary fraction and the tidal heating (not included in these
simulations) could inflate the model velocity dispersion making the
discrepancy between the MOND predictions and the data even larger.
Instead, for a lower $M/L$=0.747 ratio the predictions of MOND models
agree with the measured velocity dispersion (see Fig. \ref{efe4}). So
we conclude that the adoption of a low $M/L$ ratio, in association
with the adoption of a 5$\sigma$ rejection criterion and/or a large
orbital eccentricity, could reconcile MOND with the observed
properties of Pal 14.

\subsection{Binary fraction}
\label{binres_sec}

\begin{figure}
\epsscale{1.2}
\plotone{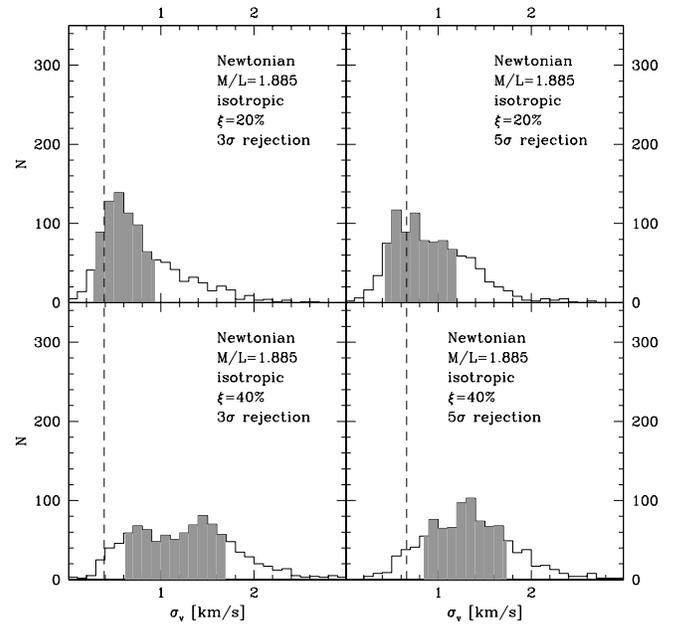}
\caption{Distribution of predicted velocity dispersions for isotropic
  Newtonian models with $f_{b}=20\%$ (upper panels) and $f_{b}=40\%$
  (bottom panels), assuming $M/L$=1.885.  Left panels refer to the
  3$\sigma$ rejection threshold, right panels to the 5$\sigma$
  rejection threshold. Grey areas indicate the range (centered on the median) comprising 68\% of
  realizations. The observed velocity dispersions are 
  marked in all panels with dashed lines.}
\label{bin1}
\end{figure}

\begin{figure*}
\epsscale{1.}
\plotone{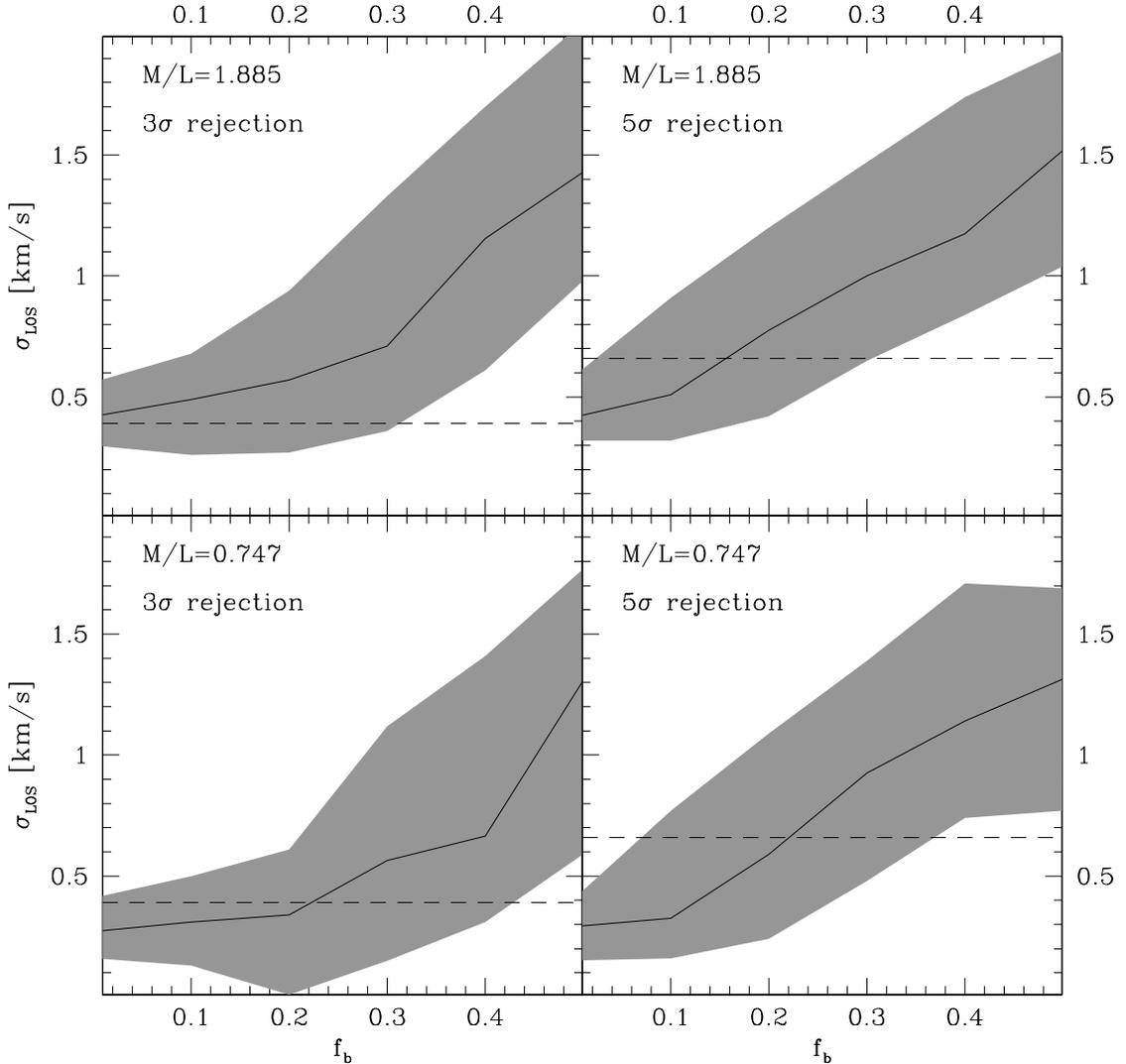}
\caption{Median (solid line) and range including 68\% of realizations
  (grey shaded area) for isotropic Newtonian models with $M/L$=1.885
  as functions of the adopted binary fraction. Left panels refer to
  the 3$\sigma$ rejection threshold, right panels to the 5$\sigma$
  rejection threshold. Upper panels refers to M/L=1.885, bottom panels to
  M/L=0.747. The observed velocity dispersions are marked in
  all panels with dashed lines.}
\label{bin2}
\end{figure*}

In Fig. \ref{bin1} the distributions of predicted velocity dispersions
are shown for (isolated) isotropic Newtonian models with $M/L$=1.885
and two different binary fractions ($f_{b}=20\%$ and $f_{b}=40\%$). As
expected, as the fraction of binaries increases the distribution
appears to be shifted to higher velocity dispersions. In
Fig. \ref{bin2} the median of the distribution and the range including
68\% of realizations are plotted as functions of the adopted binary
fraction for the two considered M/L ratios. From the comparison with the 
observed velocity dispersion of
the sample of J09 we conclude that models with $f_{b}<30\%$ in the case of 
M/L=1.885, and slightly larger $f_{b}<40\%$ for M/L=0.747 are still
acceptable.

These upper limits are larger than that derived by Kupper \& Kroupa
(2010; $f_{b}\leq 10\%$). There are three noticeable differences between
the analysis presented here and that performed by these authors: {\it
  i)} they adopt a different rejection criterion which rejects stars
outside a fixed velocity window, {\it ii)} they adopt different
characteristics for their binary population and {\it iii)} they
compute the velocity dispersion from subsamples of stars selected
randomly across the cluster. We note that while the first two choices
are somewhat arbitrary, the latter represents a limitation of Kupper
\& Kroupa (2010) analysis.  Such difference can well be responsible
for the higher predicted velocity dispersion and the corresponding smaller
limit to the cluster binary fraction found by Kupper \& Kroupa (2010).

\subsection{The effect of tidal heating in the J09 sample}

\begin{figure}
\epsscale{1.2}
\plotone{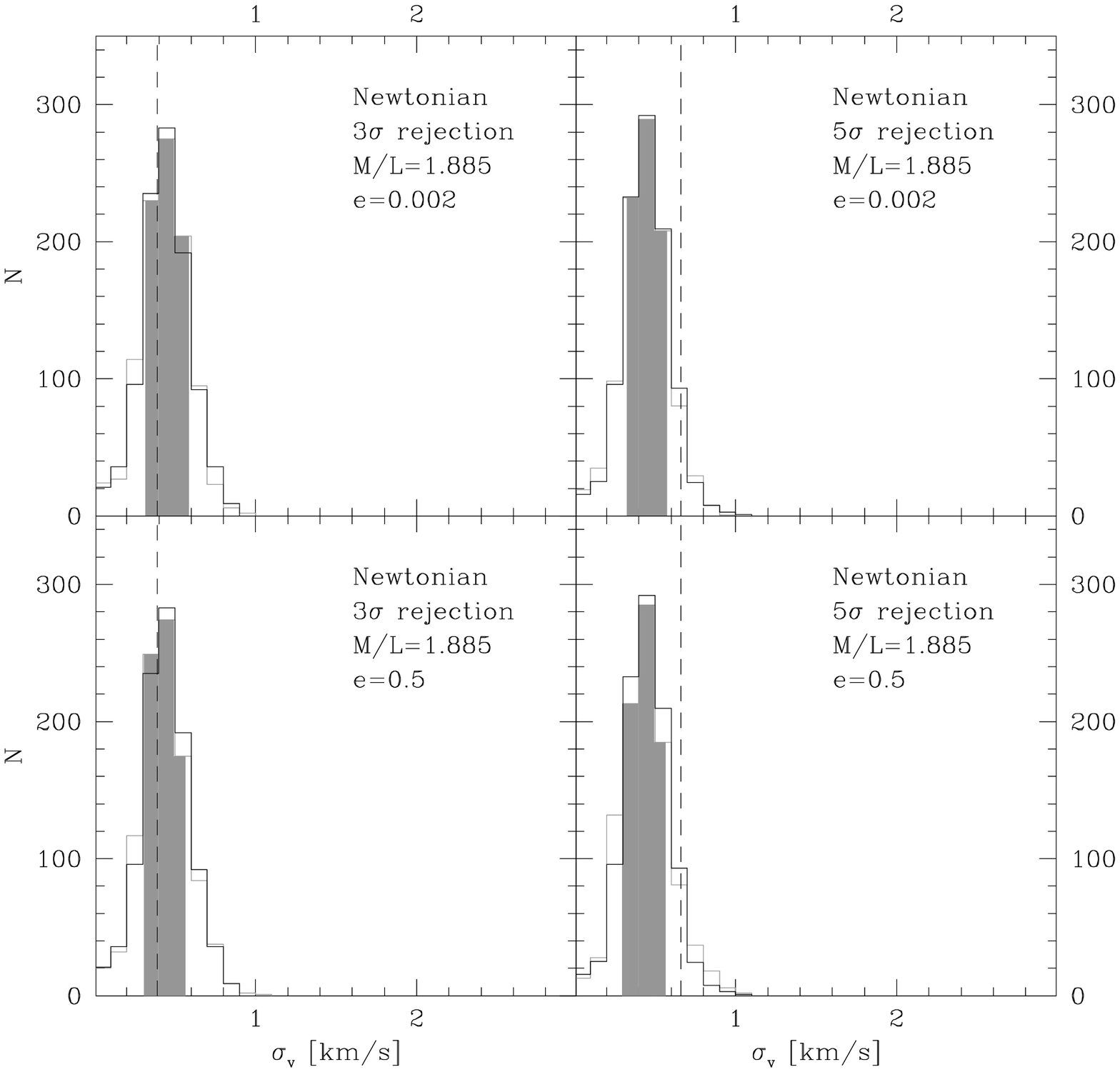}
\caption{Distribution of predicted velocity dispersions for isolated
  isotropic Newtonian models with $M/L$=1.885 and $f_{b}=0$ (black
  histograms) and models immersed in the external Galactic tidal field
  (grey histograms). In the upper and bottom panels models on
  quasi-circular orbits ($e=0.002$) and on eccentric ($e=0.5$) orbits are
  considered, respectively.  Left panels refer to the 3$\sigma$
  rejection threshold, right panels to the 5$\sigma$ rejection
  threshold. Shaded areas indicate the range (centered on the median) comprising 68\% of realizations. 
  The observed velocity dispersions are marked in all
  panels with dashed lines.}
\label{tides2}
\end{figure}

\begin{figure}
\epsscale{1.2}
\plotone{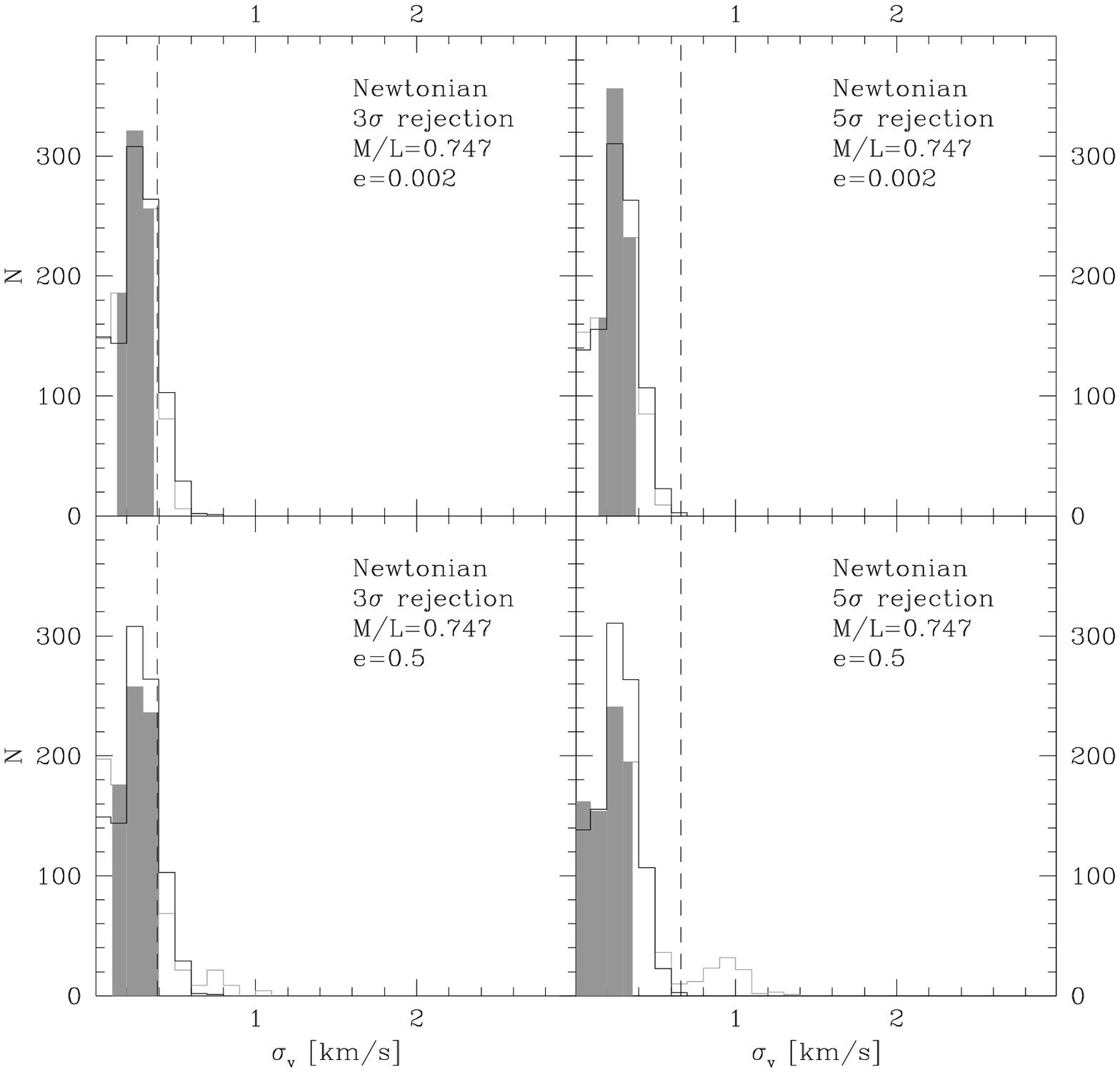}
\caption{Same as Fig. \ref{tides2}, but for models with $M/L$=0.747.}
\label{tides3}
\end{figure}

The N-body simulations described in Sect. \ref{newt_sec} allow to
evaluate the effect that tides have in the stellar kinematics
of Pal 14 in the framework of the classical Newtonian
dynamics. Quantifying this effect is important since it can
potentially affect the cluster velocity dispersion measured through
the J09 sample (Kupper et al. 2010). 

The output distributions of predicted velocity dispersions for the two
orbits are compared to the isolated model in Fig.~\ref{tides2} and
Fig.~\ref{tides3} for M/L=1.885 and M/L=0.747, respectively. It is
apparent that the effect of tidal heating is negligible in all cases.
A small difference is noticeable only in the M/L=0.747 case when
  eccentric orbits are considered: in this case the distribution,
  while having the same peak value, present a larger dispersion and a
  tail extending toward larger velocity dispersions.  The reason of
this result can be found by looking at the spatial distribution of the
J09 sample: 16 out 17 target stars reside in the inner 2 arcminutes, a
region where tides have only a minor heating efficiency (see
Fig. \ref{tides1}). The effect is indeed slightly more evident
  when a small M/L and an eccentric orbit are considered, since the
  effect of heating penetrates deeper into the cluster affecting the
  velocity of the two outermost targets. We conclude that the upper
limit in the binary fraction derived in Sect. \ref{binres_sec} is not
altered by this effect, at least for the two considered orbits.

\section{Discussion}

Using a Monte Carlo approach, we performed an accurate comparison of
the velocity dispersion of the globular cluster Pal 14 measured from
high-resolution radial velocities with the predictions of a set of
dynamical models spanning a wide range in $M/L$ ratio, degree of
anisotropy, binary fraction and orbital eccentricity in both Newtonian
and MOND gravity. The obtained results indicate that Newtonian models
with a binary fraction $f_{b}<30\%$ and a $M/L$ ratio compatible with
the predictions of stellar evolutionary models are in good agreement
with the kinematics of this stellar system. On the other hand MOND
models with the same $M/L$ ratio appear to systematically overpredict
the velocity dispersion for any assumption of the cluster
anisotropy. The same conclusion has been reached previously by J09 but
questioned by Gentile et al. (2010). Note that all these previous
approaches compared the velocity dispersion with the prediction of
theoretical models neglecting the information on the radial
distribution of targets. This has an important impact in increasing
the significance of the comparison as most of targets are actually
located in the central part of the cluster where the difference
between Newtonian and MOND models is maximized. Moreover, our simulations
showed the importance of the inclusion of the external field in the determining
the velocity dispersion profile of MOND models for this cluster. This warns against using isolated MOND
  models for clusters with masses and Galactocentric distances similar
  to Pal 14 (see also Baumgardt et al. 2005, Haghi et al. 2009,
  2011). 
Similar difficulties
for MOND in explaining the internal kinematics of GCs have been
already found in previous studies on NGC2419 (Baumgardt et al. 2009;
Sollima \& Nipoti 2010; Ibata et al. 2011a,b). In this last case the effect
of the external field, which is a factor 1.4 weaker in modulus and $\sim$23
times smaller with respect to the cluster internal acceleration than in Pal 14, 
has been estimated to be negligible (Ibata et al. 2011a).

The conclusions are different when significantly lower values of $M/L$
are considered for Pal 14. In this case the predictions of MOND
  models are close the observed velocity dispersion (in particular
  when a 5$\sigma$ rejection criterion is adopted and eccentric orbits
  are considered), while Newtonian models predict a lower velocity
  dispersion. So, a combination of small $M/L$ ratio, small binary
  fraction, relatively large orbital eccentricity and the adoption of
  a permissive rejection criterion provides a way-out for MOND.  As
  the $M/L$ ratio is a crucial parameter in the presented analysis, it
  is interesting to discuss in more detail the possibility of
  constraining its value.

From a theoretical point of view, the M/L ratio of an old, metal-poor
stellar population composed mainly by subsolar-mass stars is expected
to be systematically larger than unity ($1.5<M/L<2.5$; Fioc \&
Rocca-Volmerange 1997; Bruzual \& Charlot 2003) and can reach $M/L>3$
if dark remnants are taken into account (Kruijssen 2009). On the other
hand, observational analyses performed on GCs in the Milky Way and M31
have shown the occurrence of a sparse number of clusters with M/L
values as small as M/L$\sim$0.6 (Strader et al. 2009, 2011). It
  is worth noting that the low M/L=0.747 ratio adopted here
  corresponds to the minimum mass calculated by J09 assuming a
  decreasing mass function for stellar masses smaller than the
  limiting magnitude of deep photometric HST observations of Pal 14
  and neglecting the effect of dark remnants. So $M/L=0.747$ must be
  considered a strong lower limit.  On the other hand, while the value of 
  M/L=1.885 derived by McLaughlin \& van der Marel (2005) assumes a standard 
  Chabrier (2003) IMF, J09 measured a shallower mass
  function slope in the limited mass range covered by their observations which
  could in turn suggest a lower M/L as appropriate. In general,
  the dynamical M/L ratios measured in
  GCs appear generally smaller than those predicted by population
  synthesis models ($(M/L)_{dyn}/(M/L)_{syn}=0.82\pm0.07$ with some
  cluster with even smaller values; McLaughlin \& van der Marel 2005).
  Furthermore, in stellar systems with a small binding energy like Pal
  14, the recoil speed of white dwarf remnants could exceed the
  cluster escape velocity, causing a substantial loss of dark remnants
  (Fellhauer et al. 2003). Primordial mass segregation can also play a
  role favoring the losses of low-mass stars (with high individual M/L
  ratio; Zoonozi et al. 2011). Therefore, mass-to-light ratios
substantially lower than that predicted by stellar evolution models cannot
be excluded.  A very low binary fraction also appears disfavored (see
below), although direct estimates of the binary fraction in this
cluster are still missing.

The comparison between the observed velocity dispersion and the
Newtonian predictions sets an upper limit for the fraction of binaries
in this cluster $f_{b}<30\%$ ($f_{b}<40\%$ for the case of low
  M/L=0.747), significantly higher than the upper limit
($f_{b}<10\%$) estimated by Kupper \& Kroupa (2010; see the discussion in
Section~\ref{binres_sec}). These upper limits must be compared with
independent estimates of the binary fraction of Pal 14.
Unfortunately, the deepest photometric studies on this cluster (Dotter
et al. 2008) barely reach $\sim 3$ magnitudes below the cluster
turn-off, therefore preventing any direct estimate of the binary
fraction. However, an indirect estimate of the binary content of Pal
14 $f_{b}\sim30-40\%$ has been recently provided by Beccari et
al. (2011) on the basis of the comparison between the fraction of Blue
Straggler Stars and binary fraction in clusters with similar mass and
density.  So, on the basis of the above estimate we conclude that the
observed velocity dispersion of this cluster is consistent with the
Newtonian theory of gravity. Also in this case, this result has been
validated only for orbits with eccentricities $e<0.5$. Orbits with a
larger eccentricity can in principle lead to a significant heating of
the cluster also in its innermost region, where most of the J09
targets are located.  However, we suggest that the available data do
not provide any significant indication of a failure of the Newtonian
theory of gravity.

The present analysis can also be used to gain insight into the
  question of the dark matter content of Pal 14: in the framework of
the classical Newtonian dynamics the low observed velocity dispersion
implies a dynamical $M/L\leq 1.5-2$ (for reasonable assumptions on the
binary fraction and on the rejection criterion). This result, combined
with the observation that Pal 14 has significant tidal tails (S11),
suggests that dark matter does not contribute substantially to the
mass of this remote GC.

\acknowledgments 

We warmly thank Andreas Kupper for the helpful discussion and for
providing his data. We also thank the anonymous referee for his/her
comments and suggestions. A.S is supported by the Istituto Nazionale
di Astrofisica (INAF). C.N. is supported by the MIUR grant PRIN2008.
We acknowledge the CINECA Awards N. HP10C2TBYB (2011) and
N. HP10CQFATD (2011) for the availability of high performance
computing resources and support. \\

{\it Facilities:} \facility{CFHT}

\end{document}